 \documentclass[preprint2]{aastex}

\usepackage{epsfig}
\usepackage{epstopdf}

\usepackage{amssymb}

\shortauthors{Zagury \& Turner}
\shorttitle{The value of $R_V$}

\begin{document}

\title{The Extinction Curve in the Visible and the Value of $R_V$. \\II. Addendum to AN 333, 160}

\author{F.~Zagury$^{1}$ and D.~G.~Turner$^2$} 

\affil{
$^1$ Department of the History of Science, Harvard University, Cambridge, MA 02138, U.S.A.\\
$^2$ Department of Astronomy and Physics, Saint Mary's University, Halifax, Nova Scotia B3H 3C3, Canada}

\received{2012 April}


\begin{abstract}
This paper corrects and completes a previous study of the shape of the extinction curve in the visible and the value of $R_V$. A continuous visible/infrared extinction law proportional to $1/\lambda^p$ with $p$ close to 1 ($\pm0.4)$ is indistinguishable from a perfectly linear law ($p=1$) in the visible within observational precision, but the shape of the curve in the infrared can be substantially modified. Values of {\it p} slightly larger than 1 would account for the increase of extinction (compared to the p = 1 law) reported for $\lambda > 1 \mu$m and deeply affect the value of $R_V$. In the absence of gray extinction $R_V$ must be 4.04 if $p=1$. It becomes 3.14 for $p=1.25$, 3.00 for $p=1.30$, and 2.76 for $p=1.40$. Values of {\it p} near 1.3 are also attributed to extinction by atmospheric aerosols, which indicates that both phenomena may be governed by similar particle size distributions.
 A power extinction law may harmonize visible and infrared data into a single, continuous, and universal, interstellar extinction law. 
 \end{abstract}

\section{Introduction}

\begin{figure}[h]
\resizebox{1.\columnwidth}{!}{\includegraphics{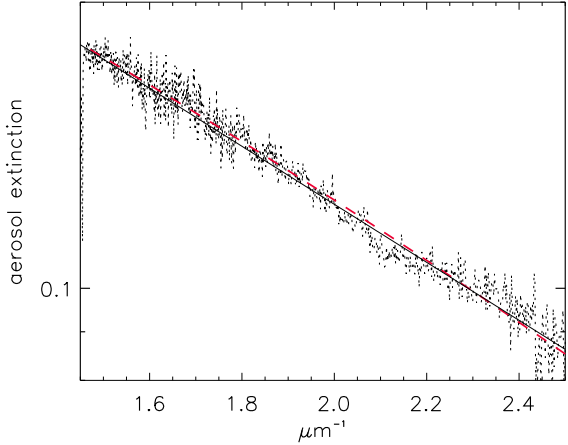}} 
\caption{The atmospheric aerosol extinction law according to an occultation spectrum of Sirius (dots) observed by the GOMOS satellite (Fig.~1 in Paper~1), after correction for ozone absorption and Rayleigh scattering by nitrogen (N$_2$). The observations are matched closely by either a linear extinction law (solid line $\propto e^{-0.7\lambda^{-1}}$) or a quasi-linear one (red dashes $\propto e^{-0.45\lambda^{-1.3}}$). 
} 
\label{fig1}
\end{figure}

``The Extinction Curve in the Visible and the Value of $R_V$,'' \citep[][hereafter Paper~1]{rv} addressed several controversial issues concerning interstellar extinction in the visible and near-infrared ($\lambda<1.2\,\mu\rm m$), in particular its wavelength dependence, variability with line of sight, and the value of $R_V=A_V/${\it E(B--V)}. To within the precision of the observations the interstellar extinction law in the visible region is well defined and depends solely on the amount of interstellar matter along the line of sight, as given by {\it E(B--V)}. Normalized extinction curves in different directions are closely reproduced by a linear extinction law over the 0.43--1.2~$\mu\rm m$ wavelength interval. In the absence of gray extinction a linear extinction law implies $R_V\sim4$, a value   \cite{tu12} found  in Carina.

As outlined in Paper~1 atmospheric aerosol extinction, much like extinction by interstellar grains, closely follows a linear extinction law (Fig.~1 in Paper~1) suggesting a close similarity in the particle size distribution for the two extinction processes. Power laws for extinction proportional to $1/\lambda^p$ result from the specific size distribution of the particles responsible for the extinction without regard for the composition of the particles.

Atmospheric aerosol extinction does not necessarily follow a simple linear law ($p=1$). According to a series of studies in the 1950s to 1970s a power law  with $p$ of order 1.3, related to a power law size distribution of particles with exponent 4.3, would best represent extinction by aerosols \citep{angstrom61, shaw73,allen73}. Fig.~\ref{fig1} demonstrates that both laws, a linear extinction law with $p=1$ and a quasi-linear one with $p=1.3$, are difficult to distinguish over the visible wavelength range alone. They are more readily separated if the spectral domain of observation is extended to the infrared. There, a quasi-linear law with $p>1$ differs markedly from a linear law (\S\ref{pec}) and displays a similar degree of flattening to what is observed in interstellar extinction observations (\S\ref{ecir}). {\it{p}}-values larger than 1 would also deeply modify $R_V$ and justify  standard $R_V$-values close to $3 $ \citep[see][]{tu76}. 

Power extinction laws ($\propto 1/\lambda^p$, $p \neq 1$) are thus less restrictive and may provide an alternative to the linear law, maybe in even better agreement with observation.
Our purpose is to investigate their properties and implications.
The study is limited to visible and infrared wavelengths extending to the $L$ band (3.5~$\mu\rm m$), beyond which thermal emission from circumstellar dust may contaminate observed extinction curves.
  
\section{Power law extinction curves}\label{pec}

\begin{figure}[h]
\resizebox{1.\columnwidth}{!}{\includegraphics{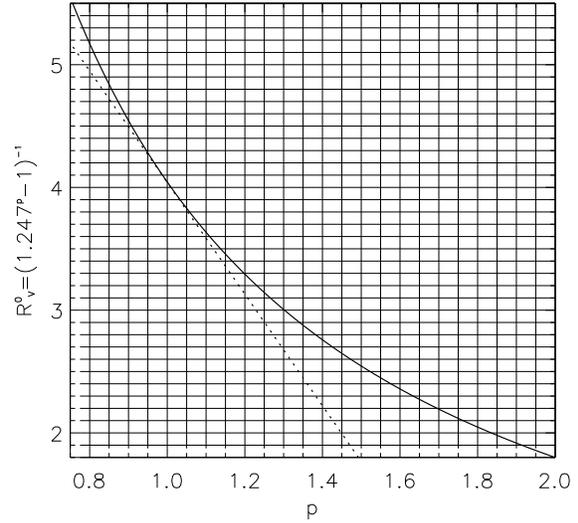}} 
\caption{The relationship between $R^{0}_V$ and $p$ for a $\lambda^{-p}$ extinction law, and the tangent at ($p=1$, $R_V=4.04$) with a slope of 4.56.} 
\label{fig2}
\end{figure}

As in \S2 of Paper~1, extinction laws $E(\lambda-\lambda_0)$ for different lines of sight, where $\lambda_0$ is a reference wavelength, are established directly from observations of reddened stars of known spectral type.
They are normalized by $E(\lambda_1-\lambda_0)$, where $\lambda_1$ is a second reference wavelength, in order to compare the extinction in different directions. 
The normalized extinction curve in the ($\lambda_0$, $\lambda_1$) system for any given direction is therefore (Eqs.~2 and 3 in Paper~1)
\begin{eqnarray}
g_{\lambda_1,\lambda_0}(x_\lambda=1/\lambda)&\,=\,&\frac{E(\lambda-\lambda_0)}{E(\lambda_1-\lambda_0)} \label{eq:ggene1}  \\
&\,=\,&\frac{A_{\lambda}}{E(\lambda_1-\lambda_0)}-R_{\lambda_1,\lambda_0}
 \label{eq:ggene2}
\end{eqnarray}
with
\begin{equation}
R_{\lambda_1,\lambda_0}=\frac{A_{\lambda_0}}{E(\lambda_1-\lambda_0)}
 \label{eq:rgene}
\end{equation}
$g_{\lambda_1,\lambda_0}(1/\lambda)$ is a measure of the wavelength-dependent extinction per unit reddening along the line of sight.

In a wavelength range where the extinction law has the form $A_\lambda\propto (x^{p}_\lambda+c_0)$, with constant $c_0>0$, that is if it is the sum of a power law and gray extinction, then 
\begin{eqnarray}
g_{\lambda_1,\lambda_0}(1/\lambda)&\,=\,&\frac{x_{\lambda}^p-x^{p}_{\lambda_0}}{x^p_{\lambda_1}-x_{\lambda_0}^{p}} \label{eq:gp}   \\
&\,=\,&\frac{1}{x^p_{\lambda_1}-x_{\lambda_0}^{p}}x_{\lambda}^p - R^{0}_{\lambda_1,\lambda_0},
 \label{eq:gpr}
\end{eqnarray}
with
\begin{equation}
R^{0}_{\lambda_1,\lambda_0}=\frac{x^{p}_{\lambda_0}}{x^p_{\lambda_1}-x_{\lambda_0}^{p}}
=\frac{1}{\left(\frac{x_{\lambda_1}}{x_{\lambda_0}}\right)^p-1}
 \label{eq:rp0}
\end{equation}
$R^{0}_{\lambda_1,\lambda_0}=g_{\lambda_1,\lambda_0}(0)$ is, by Eq.~\ref{eq:rgene},  the value of $R_{\lambda_1,\lambda_0}$ for the power law alone.
$R_{\lambda_1,\lambda_0}$ is equal to $R^{0}_{\lambda_1,\lambda_0}$ if there is no gray extinction ($c_0=0$).
It is larger otherwise.
 
In the {\it UBV} system \citep{johnson53}, $x_{\lambda_0}=1.82\,\mu\rm m^{-1}$, $x_{\lambda_1}=2.27\,\mu\rm m^{-1}$, 
\begin{equation}
g_0(x_\lambda)\,=\,\frac{x_{\lambda}^p-1.82^p}{2.27^p-1.82^p}    \label{eq:gpj} 
\end{equation}
and
\begin{equation}
R^{0}_V \approx \frac{1}{1.25^p-1}  \label{eq:rvp}
\end{equation}
Reciprocally
\begin{equation}
p=4.48\ln\left(1+\frac{1}{R^{0}_V}\right)  \label{eq:p}
\end{equation}
For no gray extinction, $c_0=0,\,R_V=R^{0}_V$, and
\begin{eqnarray}
\frac{ A_\lambda}{E\left(B-V\right)} &\,=\,& \frac{x_{\lambda}^p}{2.27^p-1.82^p}  \label{eq:ap0}\\
&\,=\,& R_V\left(\frac{x_{\lambda}}{1.82}\right)^p  \label{eq:ap}
\end{eqnarray}
For values of {\it p} near 1, $4-R^{0}_V \sim 4.56(p-1)$ from Eqs.~\ref{eq:rvp} or \ref{eq:p}. Small deviations from a linear extinction law ($p=1$) induce comparatively high variations in $R^{0}_V$.

Fig.~\ref{fig2} is a plot of the $R^{0}_V$-$ p$ relationship. In Fig.~\ref{fig3} $g_0$ functions   for different values of {\it p}, from $p=0.75$ ($R^{0}_V=5.55$) to $p=1.8$ ($R^{0}_V=2.05$), are plotted as solid curves. When $|p-1|$ increases, $g_0$ deviates more rapidly in the infrared than it does in the visible. The tangent of $g_0$ at $x_\lambda=0$ is the {\it y}-axis if $p<1$, and is parallel to the {\it x}-axis if $p>1$. The convexity at long wavelengths of a power law extinction curve is thus an indication of whether $p$ is equal to, larger than, or smaller than 1.

\section{The interstellar extinction curve in the visible}\label{pl}

Paper~1 concluded that, within observational precision, normalized extinction laws in the near-infrared/visible, from $1.2\,\mu\rm m$ to $0.43\,\mu\rm m$ ($0.8\,\mu\rm m^{-1}$ to $2.3\,\mu\rm m^{-1}$), appear independent of direction. The spectral study by \citet{divan54} or the spectrophotometric measures of \citet{nandy64,nandy65} for example, provide detailed normalized extinction curves indicating that the visible/near-infrared extinction law is linear in $1/\lambda$ to a very close approximation. The extinction in any direction over this wavelength region depends only on the amount of interstellar matter along the line of sight, as measured by {\it E(B--V)}.

In Nandy's normalization, $x_{\lambda_0}=1.22\,\mu\rm m^{-1}$ and $x_{\lambda_1}=2.22\,\mu\rm m^{-1}$, and normalized extinction curves have a slope of 1. Interstellar extinction in the visible \citep[Fig.~14 in][]{nandy64}  follows the relation
\begin{equation}
g_{Ny}(1/\lambda)=x_\lambda-1.22
 \label{eq:nec}
\end{equation}
This extinction law in conjunction with Eq.~4 of Paper~1, was used to convert Nandy's data in Cygnus \citep{nandy64} and Perseus \citep{nandy65} to the Johnson system (Fig.~\ref{fig3}). In that system the linear  law $g_{0,l}$,  dashed black line in Fig.~\ref{fig3} ($p=1$), verifies (Paper~1)
\begin{eqnarray}
g_{0,l}(\lambda)&\,=\,& \frac{2.22\,\mu{\rm m}}{\lambda}-4.04 \label{eq:g0l} \\
\frac{ A_\lambda}{E\left(B-V\right)} &\,=\,& 2.22\left[\frac{1\,\mu{\rm m}}{\lambda}+0.45\left(R_V-4.04\right)\right] \label{ajl}\\
 R^{0,l}_V&\,=\,&4.04 \label{r4}
\end{eqnarray}
In the absence of gray extinction $R_V= 4.04$ and 
\begin{equation}
\frac{ A_{\lambda,l}}{E\left(B-V\right)} \,=\, \frac{2.22\,\mu{\rm m}}{\lambda}
 \label{eq:ec}
\end{equation}
Similar to the result for aerosols (Fig.~\ref{fig1}) power laws other than Eq.~\ref{eq:g0l} fit Nandy's data  provided that the exponent {\it p} remains close to 1.

For a $g_0$ law within the limits of the two red curves in Fig.~\ref{fig3} the differences between Nandy's data and the $g_0$ curve are roughly the same as those relative to a linear extinction law ($p=1$) and agree with the mean standard error of Nandy's observations, $\pm0.05$ (in the Johnson system). The difference increases as $|p-1|$ increases and more quickly for the small number of Nandy's points under $1.5\,\mu\rm m^{-1}$. 

As $|p-1|$ increases the difference between theoretical curves with $p\neq 1$ and the linear curve $g_{0,l}$ are far more important in the infrared than in the visible (Fig.~\ref{fig3}). Infrared wavelengths therefore provide a good test of different power laws for interstellar extinction, if the same law holds in the visible and infrared.  

\section{The extinction law in the infrared from IJHKL photometry}\label{ecir}

\begin{figure} [h]
\resizebox{1.\columnwidth}{!}{\includegraphics{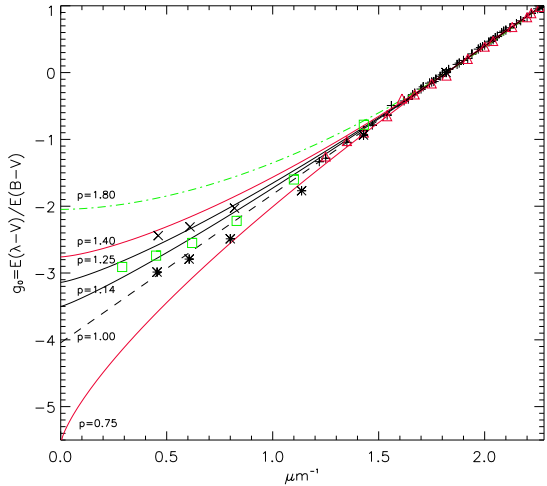}} 
\caption{Observed extinction data and calculated power laws.
 Nandy's data for Cygnus (+) and Perseus (red $\triangle$) in the visible; RIJHK,  RIJHKL, JHK photometry from \citet{berdnikov96} ($\ast$), \citet{rieke85} (green squares), and \citet{bondar06} ($\times$) in the infrared (values in Table~\ref{tbl:t1}). The power law extinction curves from bottom to top correspond to $p=0.75$ ($R_V=5.55$); $p=1.00$ ($R_V=4.04$, dashes); $p=1.14$ ($R_V=3.50$); $p=1.25$ ($R_V=3.15$);  $p=1.40$ ($R_V=2.76$); and $p=1.80$ ($R_V=2.05$).} 
\label{fig3}
\end{figure}

As was the case for the visible region there is some consensus on a unique (independent of direction) extinction law in the infrared (except perhaps in dense molecular clouds \citep{rieke85}). \citet{mathis90} concluded that ``{\it $A(\lambda)/A(J)$ is the same for $\lambda>0.9\,\mu\rm m$ for all lines of sight, to within present errors}''. Similar opinions were expressed by \citet*{berdnikov96}, \citet{sneden78}, \citet{bondar06}, and \citet{stead09}. 

Infrared normalized extinction curves published by \citet{rieke85}, \citet{berdnikov96}, and \citet{bondar06} are plotted  in Fig.~\ref{fig3}. The \citet[][green squares]{rieke85} curve, close to that of \citet{schultz75}, is often considered as a standard. All three data sets have cited uncertainties of less than $\pm0.06$, much smaller than the differences from one curve to another and therefore likely underestimated. Calibration differences may be a problem given that the curves are of similar shape despite offsets from one another. The difficulties of making ground-based observations in the infrared are also well known and arise from variable amounts of atmospheric H$_2$O absorption in the infrared. Standardization differences may ultimately be responsible for the scatter in published infrared extinction curves plotted in Fig.~\ref{fig3}, making it difficult to establish the precise run of interstellar extinction in the infrared.

Within such restrictions the  infrared data of Fig.~\ref{fig3} remain between the two red curves, consistent with a continuous power law from the visible to the infrared. Since they show an increase in interstellar extinction in the infrared and suggest a null slope at $x_\lambda=0$, the exponent of the extinction law must be greater than unity (Sect.~\ref{pec}).
It probably lies in the range $1.1 \le p \le 1.4$. The standard value of $R_V$ is then between 2.7 and 3.6 as found in many previous studies \citep[e.g.][]{tu76}. Exponents {\it p} larger than 1.8 which have been suggested  by \citet{mathis90}, \citet{berdnikov96}, or \citet{stead09}, cannot be reconciled with the assumption of a universal extinction law from the visible to the infrared. Either the near-infrared extinction law differs from that for the visible, in which case the extrapolation of the extinction curve to $1/\lambda=0$ does not give $R_V$ (Paper~1), or the same power law is valid for both the visible and infrared normalized extinction curves, and an updated version of the infrared extinction law is needed in order to fix the exact values of {\it p} and $R_V$.
 
\setcounter{table}{0}
\begin{table*}
\begin{center}
\caption[]{Near-infrared photometry}		
\begin{tabular}{lcccccc}
\hline
 & \multicolumn{5}{c}{Johnson Photometry}  \\
band&  R&I&J&H&K&L \\
\hline
wavelength  ($\mu$m) & 0.70& 0.91& 1.20& 1.61 & 2.22& 3.45\\
wavenumber ($\mu\rm m^{-1}$) & 1.43&1.1&0.83&0.62&0.45&0.29 \\
\hline
& \multicolumn{5}{c}{Interstellar extinction normalized colors}  \\
Source$^{(1)}$ &$\frac{E(R-V)}{E(B-V)}$ &$\frac{E(I-V)}{E(B-V)}$ &$\frac{E(J-V)}{E(B-V)}$ &$\frac{E(H-V)}{E(B-V)}$ &$\frac{E(K-V)}{E(B-V)}$ &$\frac{E(L-V)}{E(B-V)}$\\
\hline
RL85&-0.78&-1.60&-2.22&-2.55&-2.74&-2.91  \\
BVD96&-0.93& -1.77 & -2.49& -2.79& -2.99&\\
B06&&&-2.03&-2.31&-2.44&  \\
p=0.75 &-0.92&-1.74&-2.47&-3.07& -3.60&-4.15\\
p=1.00&-0.87&-1.60&     -2.20&    -2.67&     -3.04&   -3.40\\
p=1.13&-0.84&-1.53&-2.07&-2.48& -2.80&-3.08\\
p=1.25&-0.82&  -1.45& -1.96& -2.32& -2.59& -2.82\\
p=1.40&-0.79&-1.39&-1.84&-2.15& -2.37&-2.55\\
p=1.80&-0.72&-1.22&-1.55&-1.75& -1.88&-1.97\\
\hline
\end{tabular}
\end{center} 

\begin{list}{}{}
\item[$(1)$] Data from \citet[][RL85]{rieke85}), \citet[][BVD96]{berdnikov96}, \citet[][B06]{bondar06}, and from the calculated values for a power law extinction with $p=0.75$ ($R_V=5.55$), $p=1.00$ ($R_V=4.04$), $p=1.14$ ($R_V=3.50$), $p=1.25$ ($R_V=3.14$), $p=1.40$ ($R_V=2.76$) and $p=1.80$ ($R_V=2.05$).
\end{list}
\label{tbl:t1}
\end{table*}

\section{Conclusion}

In Paper~1 and here several issues are addressed of importance to a comprehensive understanding of interstellar extinction. What is the shape of the visible/near-infrared extinction curve? Does the extinction law vary with line of sight? And what is the value of $R_V$? Such questions arise from published results for the visible/near-infrared extinction curve that are incompatible with each other. It is not possible for  the value of $R_V$ (and the extinction law) to be independent of the line of sight, as suggested by different data sets, yet to vary with direction viewed, as implied in other studies \citep[][and Paper~1]{tu76,tu89,tu94}. Infrared and visible data suggest that the extinction in any direction, in the infrared/visible wavelength range, depends solely on the quantity of interstellar matter along the sight line, quantified by {\it E(B--V)}. That is consistent with a common, direction-independent, extinction law governing both wavelength domains. A similar law holds for atmospheric aerosols.

In the visible spectral region different power laws with exponents close to 1 describe equally well the observed normalized interstellar extinction curve and permit a large range of $R_V$ values. The recent study by \citet{tu12} demonstrates that uncertainties in direct estimates (from photometry in the visible) of $R_V$ can be large. Provided that extinction in the visible and infrared results from the same continuous size distribution of interstellar grains, values of {\it p} and $R_V$ are more easily established from infrared data. However, definitive conclusions about the exact wavelength dependence of the infrared interstellar extinction curve are difficult to reach. Extinction in the infrared is small and ground-based observations are limited to  a few photometric bands. Observational uncertainties are consequently large and conspicuous differences exist between published infrared extinction curves. Additional studies of interstellar extinction, presumably from space and from spectroscopic observations, covering both the visible and the infrared, would be most beneficial. 

{}
\end{document}